\documentclass[aps,prd,preprint,floatfix,nofootinbib,showpacs]{revtex4-1}
\usepackage{hyperref,amssymb}
\usepackage{graphicx,color,epstopdf,amsmath}
\newcommand{\BM}[1]{{\mbox{\boldmath{$#1$}}}}
\newcommand{\fr}[2]{{\hbox{$ #1 \over #2 $}}}

\begin{document}
\title{Leptoquark induced rare decay amplitudes 
$h \to \tau^\mp \mu^\pm$  and $\tau\to \mu\gamma$}
\author{ Kingman Cheung$^{1,2,3}$, Wai-Yee Keung$^{4,1}$, Po-Yan Tseng$^{2}$}
\affiliation{
$^1$ Physics Division, National Center for Theoretical Sciences, Hsinchu,
Taiwan \\
$^2$Department of Physics, National Tsing Hua University, Hsinchu 300, Taiwan \\
$^3$Division of Quantum Phases \& Devices, School of Physics, 
Konkuk University, Seoul 143-701, Republic of Korea \\
$^4$Department of Physics, University of Illinois at Chicago, Illinois 60607 USA \\
}

\renewcommand{\thefootnote}{\arabic{footnote}}
\date{September 30, 2015}

\begin{abstract}
Rare decay modes of the newly discovered standard-model-like Higgs
boson $h$ may test the flavor changing couplings in the leptoquark
sector through the process $h \to \tau^\mp\mu^\pm$.  Motived by the
recently reported excess in LHC data from the CMS detector, 
we found that a predicted
branching fraction Br($h \to \tau^\mp\mu^\pm$) at the level of 1\% is
possible even though the coupling parameters are subjected to the
stringent constraint from the null observation of $\tau\to\mu\gamma$,
where the destructive cancellation among amplitudes is achievable  
by fine tuning.
\end{abstract}

\maketitle

\section{Introduction}
The newly discovered Higgs boson $h$ at the mass 125 GeV is consistent
with the Higgs boson predicted in the standard model (SM) 
\cite{higgcision}. The narrow
decay width of a predicted size about 4 MeV in SM provides hope that
the unusual rare decay due to new physics (NP) can have a measurable branching
fraction. 
Recently, the CMS collaboration has reported~\cite{Khachatryan:2015kon}
a possible excess in the decay process $h\to\tau^\mp\mu^\pm$ with a
significance of 2.4 $\sigma$ in the search for the lepton flavor
violation (LFV).
Assuming SM Higgs production, CMS obtained the best fit 
for the branching fraction summed over $\tau^-\mu^+$ and $\tau^+\mu^-$,
\begin{equation}
\label{data}
\hbox{Br}(h\to\tau^\mp \mu^\pm) = 0.84^{+0.39}_{-0.37}\, \%   \ . 
\end{equation}
We understand that it is too early to draw a positive inference until
future analyses of higher statistics from both CMS and ATLAS
experiments are performed. However, the present sensitivity at the
$1\%$ level is interesting enough to call for possible NP to deliver
such a detectable rate but satisfy other rare decay constraints such
as $\tau\to\mu\gamma$,
\begin{equation}
\hbox{Br}(\tau\to\mu\gamma) < 4.4\times 10^{-8} 
\hbox{ at 90\% C.L. from BaBar experiment \cite{Aubert:2009ag} }
\end{equation}
at low energy. 
Indeed, there are a lot of
theoretical activities\cite{Dorsner:2015mja, Harnik:2012pb,Campos:2014zaa,Celis:2014roa, Sierra:2014nqa,Lee:2014rba,Heeck:2014qea,Crivellin:2015mga,deLima:2015pqa,11,Das:2015zwa,12,13,14,15,He:2015rqa,16,17,18} 
along this line of investigation. 
There were also a number of studies on LFV Higgs boson decays
in literature \cite{old}.
We are particularly motivated by the leptoquark (LQ) associated with
the third generation, which provides a large top quark mass
insertion in the loop diagram. However, the LQ interactions also give
rise to amplitudes for $\tau \to \mu \gamma$. We notice that the cancellation
between two types of LQ contributions is possible for $\tau \to \mu
\gamma$, leaving a large detectable decay rate for 
$h\to \tau^\mp \mu^\pm$. Each of these two types of LQs has been outlined in
the literature, such as in
Ref.\cite{Dorsner:2015mja}, but the combined version necessary for
the cancellation was overlooked. 

The organization of the work is as follows. In the next section, we 
describe the LQ interactions associated with the top quark and tau lepton.
In Sec. III and IV, we calculate the decay $\tau \to \mu\gamma$ and
$h \to \tau^\mp \mu^\pm$, respectively. We give details on numerical 
results in Sec. V. We conclude in Sec. VI.

\section{Leptoquark interactions associated with the top quark}

\begin{figure}[t!]
\includegraphics[width=3in]{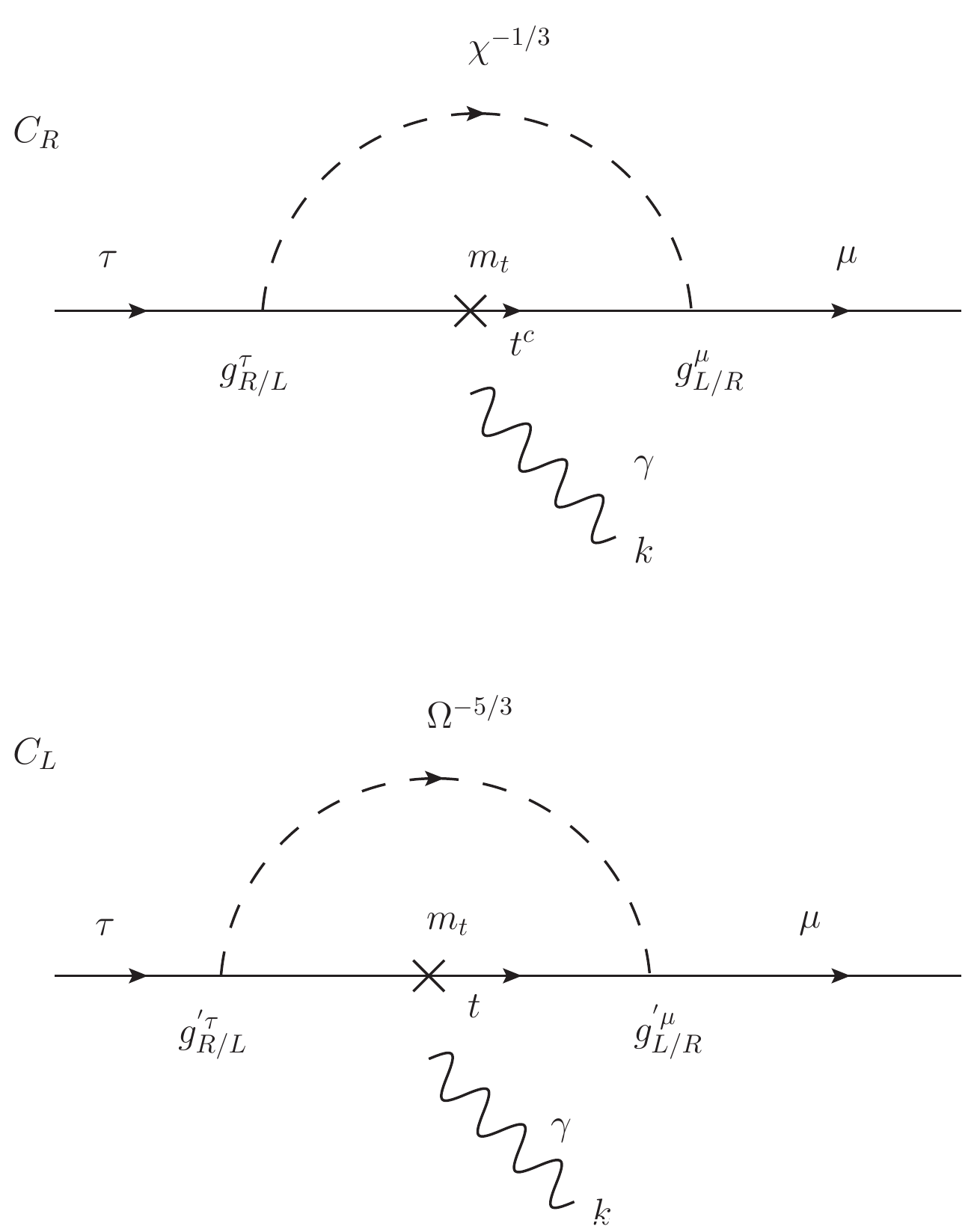}
\caption{\small \label{fig1}
(a) Dipole transition of $\tau\to\mu\gamma$ via the singlet LQ $\chi$.
(b) Dipole transition of $\tau\to\mu\gamma$ via the 
doublet LQ $\Omega$.
}
\end{figure}

We associate the new LQs with the top quark of the third generation in
order to avoid the very stringent constraints upon the flavor non-conservation 
among the first two generations. 
On the other hand, the mass insertion of the top quark
can enhance the rate of the rare LFV Higgs decay mode 
among the second and the third generation leptons. 
To satisfy the electroweak gauge symmetry, we can
classify two  types of LQs, 
the first one $\chi^{\fr13}$ is a weak $SU(2)$ singlet, 
and the other one  $SU(2)$ doublet, {\it i.e.}
$\Omega^T=(\Omega^{\fr53}, \Omega^{\fr23})$.
The superscript denotes the electromagnetic charge number.
LQs transform under the $SU(3)$ color group as $\bf 3$ just like  quarks. 
The relevant interactions are given by
\begin{eqnarray}
 {\cal L} & \supset  &
        g_L^\tau \chi^{\fr13} (Q_3)_L^T \epsilon L_{\tau,L}
        -g_R^\tau \chi^{\fr13} t_R \tau_R      \nonumber\\    
 &&  +   {g'}_L^\tau     \Omega^T \epsilon \overline{t_R} L_{\tau,L}
      - {g'}_R^\tau  \overline{Q_{3,L}} \tau_R \Omega
\  + ( \ \tau \leftrightarrow  \mu \ )   + \hbox{ h.c. }  
\end{eqnarray}
The Feynman diagrams (for $\tau \to \mu \gamma$) that involve these 
LQ interactions  are shown in Fig.~\ref{fig1}.
The shown $\epsilon$ symbol, basically $i\sigma_2$,
links two $SU(2)$ doublets into a gauge invariant singlet. 
For brevity, we do not show other Levi-Civita symbols 
that contract Weyl spinor indexes. Also,
$(Q_3)_L^T=(t_L,b_L)$, $L_{\tau,L}=(\nu_{\tau,L}, \tau_L)^T$.
The terms from exchanging $\tau \leftrightarrow  \mu$ are needed to induce the 
LFV between the muon and the tau.

\section{$\tau\to\mu\gamma$ amplitudes  induced by leptoquarks}
We start with the contributions from the singlet $\chi$.
We define $t^c$ to be the charge conjugated state of $t$. In this way,
we can avoid the use of the unfamiliar Feynman rule for two fermions 
flowing into
a vertex.  Instead, one fermion flows in and the other out.
For example, the incoming $\tau$ enters the first vertex and 
turns into a departing $t^c$  plus a boson $\chi^{-{1\over3}}$.
The relevant vertices for the process $\tau \to \mu\gamma$  are
\begin{equation}  
g_L^\tau (\chi^{-{1\over3}})^\dagger (\overline{t^c}\tau_L )
-g_R^\tau (\chi^{-{1\over3}})^\dagger (\overline{t^c}\tau_R ) + 
\ ( \ \tau \leftrightarrow  \mu \ ) \ +  \hbox{ h.c. }  
\end{equation}
For the outgoing left-handed muon, the Feynman amplitude that 
the external photon line attaches to the $t^c$ line is given by
\begin{equation}
i{\cal M}_1 (\tau \to \mu\gamma)=
-\frac{e q_{t^c}  g_R^\tau  g_L^\mu m_t}{16\pi^2} 3_c
\int_0^1 \frac{(1-z)^2 dz }{zm_\chi^2 +(1-z) m_t^2 }
\sigma^{\mu\nu}k_\nu R \;,      
\ . \end{equation}
where $R$ stands for the right-handed chiral projection operator 
$(1+\gamma^5)/2$.
It is understood that the external spinors 
$\overline{u(\mu)}$ and $u(\tau)$ sandwich the Dirac chain. 
We keep track of the color factor 3 by a subscript $c$.
Another amplitude  where the photon attaches to $\chi^{-{1\over3}}$ 
is
\begin{equation}
 i{\cal M}_2 (\tau \to \mu\gamma)=
\frac{e q_{\chi^{-{1\over3}}}  g_R^\tau  g_L^\mu  m_t}{16\pi^2} 3_c
\int_0^1 \frac{(1-z)z dz }{zm_t^2 +(1-z) m_\chi^2 }
\sigma^{\mu\nu}k_\nu R       \ .
\end{equation}
We set charges $q_{t^c}=-\fr23$ and $q_{\chi^{-{1\over3}}}=-\fr13$. 
Using $z\leftrightarrow (1-z)$ in  ${\cal M}_1$, we obtain
\begin{equation}
 i{\cal M}_{1+2}=\frac{e   g_R^\tau  g_L^\mu m_t}{16\pi^2 m_\chi^2} 3_c
\int_0^1 \frac{   \fr23 z^2 -\fr13 z(1-z)dz }{(1-z)+zm_t^2/m_\chi^2 }
\sigma^{\mu\nu}k_\nu R   
\ .\end{equation}
The numerator of the integral becomes $z^2-\fr{z}3$. The overall result is
\begin{equation}
  i{\cal M}_{1+2}=\frac{e   g_R^\tau  g_L^\mu m_t}{16\pi^2 m_\chi^2} 3_c
\left( \xi_1(x_t)-\fr13\xi_0(x_t) \right)  \sigma^{\mu\nu}k_\nu R 
\ ,\quad x_t=m_t^2/m_\chi^2 \ , 
\hbox{ and }  
\ .\end{equation}
\begin{equation}
\xi_n(x) \equiv \int^1_0 \frac{z^{n+1}{\rm d}z}{1+(x-1)z}
=-{ {\rm ln}x+(1-x)+\cdots+{(1-x)^{n+1}\over n+1} \over (1-x)^{n+2} }
\ ,    \end{equation}
$$ \quad  \hbox{ and } \xi_{-1}(x) \equiv \int^1_0 \frac{{\rm d}z}{1+(x-1)z}
=-{ {\rm ln}x\over 1-x}
\ . $$
So the amplitude is related to the integral function,
\begin{equation}
H_1(x) \equiv  \xi_1(x)-\fr13\xi_0(x)=-\frac1{6(1-x)^3}\left[
7-8x+x^2+2(2+x)\ln(x)\right]
\ . \end{equation}
Our result is different from that in Ref.\cite{Dorsner:2015mja}.
Note that there is another chiral amplitude for the outgoing
right-handed muon, using $g_L^\tau g_R^\mu$. These two amplitudes do
not interfere in the zero muon mass limit.

Now we switch to the contributions from the LQ doublet $\Omega$.
The relevant vertices for the process $\tau \to \mu\gamma$  are
\begin{equation}
 -  {g'}_R^\tau (\Omega^{\fr53}) (\overline{t_L}\tau_R )
+ {g'}_L^\mu (\Omega^{\fr53}) (\overline{t_R}\mu_L )
 \  + ( \ \tau \leftrightarrow  \mu \ )  + \hbox{ h.c. }  
\end{equation}
For the outgoing left-handed muon,
\begin{equation}
 i{\cal M'}_1 (\tau \to \mu\gamma)=
-\frac{e q_{t}  {g'}_R^\tau  {g'}_L^\mu m_t}{16\pi^2} 3_c
\int_0^1 \frac{(1-z)^2 dz }{zm_\Omega^2 +(1-z) m_t^2 }
\sigma^{\mu\nu}k_\nu R       
\ .\end{equation}
This corresponds to the diagram that the external photon line attaches to the 
$t$ line. 
Another amplitude  where the photon attaches to $\Omega^{-{5\over3}}$ 
is
\begin{equation}
 i{\cal M'}_2 (\tau \to \mu\gamma)=
\frac{e q_{\Omega^{-{5\over3}}}  {g'}_R^\tau  {g'}_L^\mu  m_t}{16\pi^2} 3_c
\int_0^1 \frac{(1-z)z dz }{zm_t^2 +(1-z) m_\Omega^2 }
\sigma^{\mu\nu}k_\nu R       
\ .\end{equation}
We set charges $q_{t}=\fr23$ and $q_{\Omega^{-{5\over3}}}=-\fr53$. 
Using $z\leftrightarrow (1-z)$ in  ${\cal M}_1$, we obtain
\begin{equation}
 i{\cal M'}_{1+2}
=\frac{e   {g'}_R^\tau  {g'}_L^\mu m_t}{16\pi^2 m_\Omega^2} 3_c
\int_0^1 \frac{   -\fr23 z^2 -\fr53 z(1-z)dz }{(1-z)+zm_t^2/m_\Omega^2 }
\sigma^{\mu\nu}k_\nu R       \ .\end{equation}   
The numerator of the integral becomes $z^2-\fr{5z}3$. The overall result is
\begin{equation}
  i{\cal M'}_{1+2}
=\frac{e   {g'}_R^\tau  {g'}_L^\mu m_t}{16\pi^2 m_\Omega^2} 3_c
\left( \xi_1(x'_t)-\fr53\xi_0(x'_t) \right) \sigma^{\mu\nu}k_\nu R   \ ,
\quad x'_t=m_t^2/m_\Omega^2 \ .\end{equation}
So the amplitude is related to the integral function,
\begin{equation}
 H_2(x) \equiv  \xi_1(x)-\fr53\xi_0(x)=-\frac1{6(1-x)^3}\left[
-1+8x-7x^2-2(2-5x)\ln(x)\right] 
  \ .
\end{equation}
Note that there is another chiral amplitude for the outgoing right-handed muon,
using ${g'}_L^\tau  {g'}_R^\mu$.

In general, the low energy effective operators of dim 5 are
\begin{equation}
{\cal L}_{\rm eff} \supset \frac{e}{m_t} 
\left[ \bar\mu \sigma^{\alpha\beta} (C_L L +  C_R R)\tau \right] 
F_{\alpha\beta} + \hbox{ h.c. } 
\end{equation}
\begin{equation} 
C_R= \frac{ 3_c}{32\pi^2 }\left( 
 g_R^\tau  g_L^\mu  x_t H_1(x_t) + {g'}_R^\tau{g'}_L^\mu x'_t H_2(x'_t) \right)
\ ,
\end{equation}
\begin{equation}
C_L= \frac{ 3_c}{32\pi^2 }\left( 
 g_L^\tau  g_R^\mu  x_t H_1(x_t) + {g'}_L^\tau{g'}_R^\mu x'_t H_2(x'_t) \right)
\ . 
\end{equation}
The partial decay width of the process $\tau\to\mu\gamma$ is
\begin{equation}
   \Gamma(\tau\to\mu\gamma)= \frac{e^2}{4\pi} m_\tau
\left(\frac{m_\tau^2}{m_t^2}\right)
(|C_L|^2+|C_R|^2)  \ . 
\end{equation}
The general loop formulas for the radiative transitions can be found in 
Ref.\cite{Lavoura:2003xp}
\section{$h \to \tau + \bar\mu$  via Leptoquarks of the 3rd generation}
\begin{figure}[t!]
\includegraphics[width=3in]{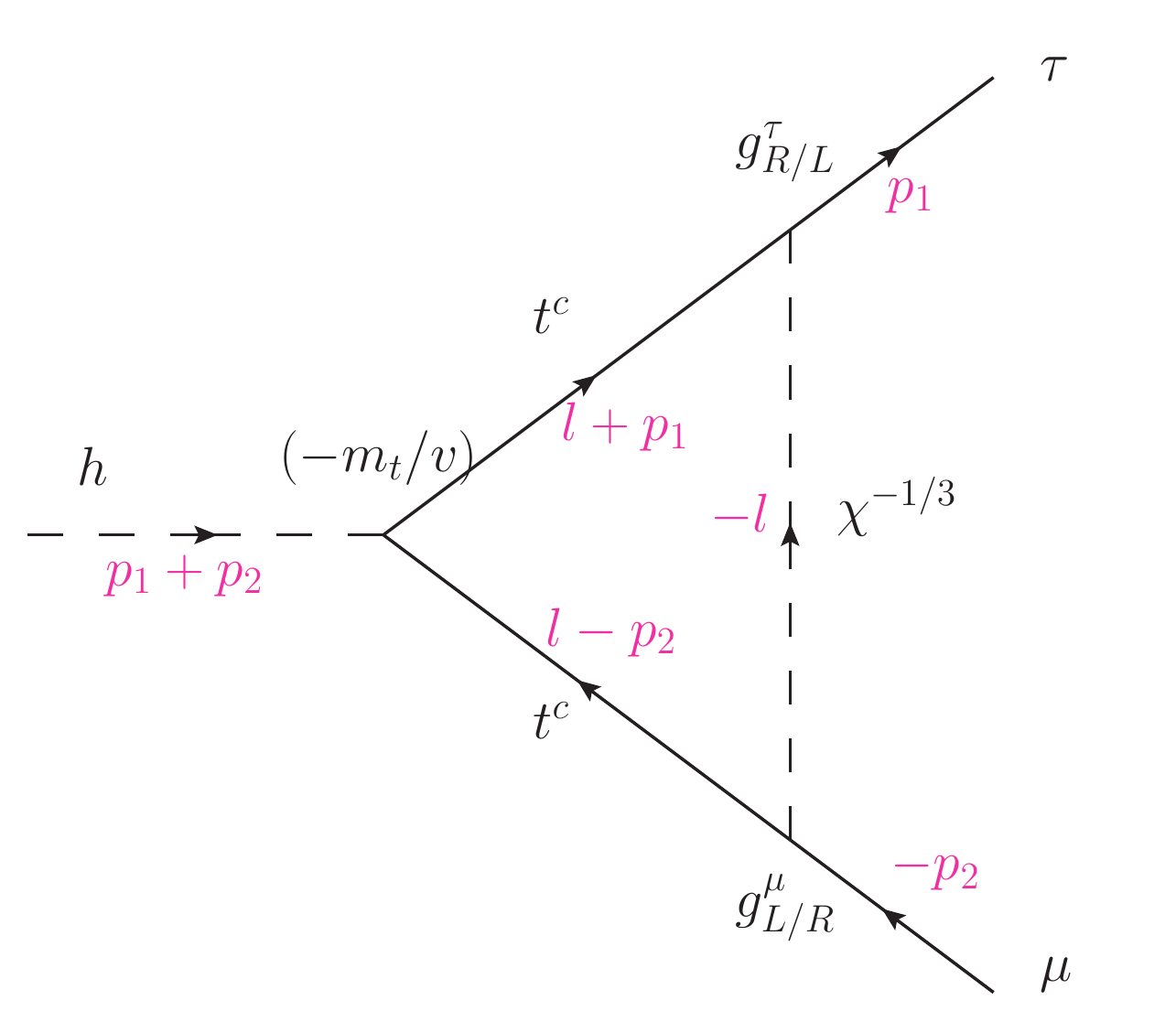}
\caption{\small \label{fig2}
$ h\to \tau\bar\mu$ via the singlet LQ $\chi$.
}
\end{figure}

For the rare decay $h\to \tau\mu$, 
we start with the contribution from
the $SU(2)$ singlet leptoquark $\chi^{-\fr13}$ to the chiral amplitude 
of the outgoing right-handed $\tau$. 
We take the zero mass limit for $\mu$ and $\tau$.
At the one loop level, the Higgs coupling to 
$\tau (p_1) \bar\mu (p_2) $ is
induced via a triangle diagram, which involves internal $t^c,\chi$ lines. 
First, we concentrate at the diagram that the external Higgs touches 
the internal $t^c$ line, as shown in Fig.~\ref{fig2}.

\begin{equation}
i{\cal M}_{\chi,R}^\triangleleft=(i)^6 \   3_c
\left(-g_R^\tau g_L^\mu \right) \int \frac{d^4\ell}{(2\pi)^4} 
 \frac{
L
(\not{\ell}+\not{p}_1+m_t) (-\fr{m_t}{v}) (\not{\ell}-\not{p}_2 + m_t)
L}
    {((\ell+p_1)^2-m_t^2) ((\ell-p_2)^2-m_t^2)(\ell^2-m_\chi^2)}   
\ .\end{equation}
We use the Feynman parameterization trick to carry over the integration. 
The parameters $\alpha,\beta,\gamma$ are assigned to  the denominator factors
$(\ell+p_1)^2-m_t^2$, $(\ell-p_2)^2-m_t^2$, $\ell^2-m_\chi^2$ respectively,
under the constraint $\alpha+\beta+\gamma=1$.
Then we complete the square of the denominator as follows,
$$ \alpha[(\ell+p_1)^2-m_t^2]+\beta[(\ell-p_2)^2-m_t^2]
+\gamma[\ell^2-m_\chi^2] $$
$$ = \ell^2 +2\alpha p_1\cdot \ell -2\beta p_2\cdot\ell+\alpha p_1^2
+\beta p_2^2 -(\alpha+\beta)m_t^2 -\gamma m_\chi^2 $$
\begin{equation}
=(\ell+\alpha p_1-\beta p_2)^2 -m^2(\alpha,\beta)
\hbox{ , where }
m^2(\alpha,\beta)=m_\chi^2-\alpha\beta s+
 (\alpha+\beta)(m_t^2-m_\chi^2)
\ .\end{equation}
Shifting the loop momentum, we simplify
the numerator of the the Dirac matrices with the equation of motion,
$$ (\not{\ell}^2+m_t^2) \longrightarrow
   \ell'^2-2\alpha\beta p_1\cdot p_2 +m_t^2     
\longrightarrow \ell'^2 -\alpha\beta s +m_t^2 
\ . $$
Here the $s$ variable is simply $2 p_1\cdot p_2 =m_h^2$.
The  amplitude becomes
\begin{equation}
 i{\cal M}_{\chi,R}^\triangleleft
=-3_c \left(  g_R^\tau g_L^\mu \fr{m_t}{v}  \right) 
\int_{{\tt L\!\!\backslash}} 2! d\alpha d\beta\int \frac{d^4\ell'}{(2\pi)^4}
\frac{\ell'^2-\alpha\beta s +m_t^2 }{[\ell'^2-m^2(\alpha,\beta)]^3} L  
\ .\end{equation}
The domain $\tt L\!\!\backslash$ covers positive 
$\alpha$ and $\beta$, as well as $\alpha+\beta\le 1$.

We perform the Wick's rotation by Euclideanizing $\ell'^0 \to iq_{E4}$,
$\ell'^2\to -q_E^2$, 
$d^4\ell'\to id^4 q_E$, 
$d^4 q_E \to d^3\BM{q}_E dq_{E4}=4\pi|\BM{q}_E|^2 d|\BM{q}_E| dq_{E4}
\to 4\pi (q_E^2\cos^2\phi)\fr12 dq_E^2 d\phi 
\to \pi^2 q_E^2 dq_E^2$. So
$$ {\cal M}_{\chi,R}^\triangleleft=-3_c
\left(  g_R^\tau g_L^\mu \fr{m_t}{v} \right)   
\int_{{\tt L\!\!\backslash}}   2! d\alpha d\beta
\int 
\frac{-q_E^2 -\alpha\beta s + m_t^2 } 
     {-[q_E^2+m^2(\alpha,\beta,s)]^3} 
\frac{q_E^2 dq_E^2}{16\pi^2} L
  \ ,$$
\begin{equation}
 \longrightarrow 
\quad -3_c  
\frac{ g_R^\tau g_L^\mu  m_t}{16\pi^2 v}   \int_{{\tt L\!\!\backslash}}\left(
\log\frac{\Lambda^2}{m^2(\alpha,\beta,s)}-\frac{3}{2}
+\frac{\alpha\beta s - m_t^2}{ 2 m^2(\alpha,\beta,s)}\right)2! d\alpha d\beta
L  \ . \end{equation}
The logarithmic divergence has to be canceled by the 
one-particle reducible (1PR) diagrams 
with bubbles in the external lepton lines.
The $h$ line is either attached directly to  $\tau$ or $\mu$, picking up 
respectively the mass couplings $m_\tau$ or $m_\mu$, which are canceled 
by the propagators.
It can be shown the corresponding 1PR contribution to be
\begin{equation}
  \longrightarrow
\quad +3_c
\frac{ g_R^\tau g_L^\mu  m_t}{16\pi^2 v}   \int_0^1
\left(\log\frac{\Lambda^2}{\gamma m_\chi^2 + (1-\gamma) m_t^2}-1\right)
 d\gamma L  \ . \end{equation}
Overall, $\log\Lambda^2$ terms cancel.
Therefore the combined  amplitude is
\begin{equation}
 {\cal M}_{\chi}=-3_c \frac{1}{16\pi^2}\frac{m_t}{v} G_\chi \ 
 ( g_R^\tau g_L^\mu  L + g_R^\mu g_L^\tau  R ) \ .\end{equation}
\begin{equation}
 G_\chi=
\int_{{\tt L\!\!\backslash}}\left( \log\frac{\Lambda^2}{m^2(\alpha,\beta,s)}
-\frac12 +\frac{\alpha\beta s - m_t^2}{ 2 m^2(\alpha,\beta,s)}
\right)2! d\alpha d\beta   
-\int_0^1  \log\frac{\Lambda^2}{\gamma m_\chi^2 + (1-\gamma) m_t^2}
   d\gamma  \ . \end{equation}
Note that in the intermediate step, we can choose 
an arbitrary $\Lambda$ for the convenience of the calculation.

Alternatively, one can use the Passarino-Veltman\cite{Passarino:1978jh}
(PV) functions.
The integral before the Feynman's parameterization as given in (21) is
\begin{eqnarray}
&  \int 
\frac{ \fr{d^4\ell}{(2\pi)^4} (\ell^2 +m_t^2)}
     {((\ell+p_1)^2-m_t^2)((\ell-p_2)^2-m_t^2)(\ell^2-m_\chi^2)}
=\int 
\frac{ [(\ell^2-m_\chi^2) \ + \ m_\chi^2+ m_t^2]\frac{d^4\ell}{(2\pi)^4}}
     {((\ell+p_1)^2-m_t^2) ((\ell-p_2)^2-m_t^2)(\ell^2-m_\chi^2)}  \nonumber\\
& = \int \frac{d^4\ell}{(2\pi)^4}\left(
\frac{m_\chi^2+ m_t^2}     
      {((\ell+p_1)^2-m_t^2) ((\ell-p_2)^2-m_t^2)(\ell^2-m_\chi^2)}  
+\frac{1} {((\ell+p_1)^2-m_t^2) ((\ell-p_2)^2-m_t^2)}
\right) \ .     \end{eqnarray}
The first term  simply gives the triangle function 
\begin{equation}
 \fr{i}{16\pi^2}(m_t^2+m_\chi^2) C_0(0,0,s,m_t^2,m_\chi^2, m_t^2) \ .
\end{equation}
The second term after shifting the loop momentum gives  the bubble function
\begin{eqnarray}
\fr{i}{16\pi^2}B_0(s,m_t^2,m_t^2) &=\int \frac{ {d^4\ell}/{(2\pi)^4}} 
          {((\ell+p_1)^2-m_t^2) ((\ell-p_2)^2-m_t^2)}   
          =\int \frac{       {d^4\ell'}/{(2\pi)^4}}
          {((\ell'+p_1+p_2)^2-m_t^2) (\ell'^2-m_t^2)}
\ . \end{eqnarray}
The result including the 1PR bubbles is
\begin{equation}
G_\chi = (m_\chi^2+m_t^2) C_0(0,0,s,m_t^2,m_\chi^2,m_t^2)
           +B_0(s,m_t^2,m_t^2)  \ - B_0(0,m_t^2,m_\chi^2) 
   \ .\end{equation}
We have cross-checked numerically that the $G$ value from PV
functions and from the Feynman parameterization method match each other.

For the Feynman diagram that the Higgs touches the leptoquark, the
required vertex originates from the bosonic interaction of 
$-\lambda_\chi H^\dagger H \chi^\dagger \chi$. 
The $G$ coefficient is updated with 
the new addition,
$$  G_\chi \to G_\chi + \lambda_\chi v^2 \, C_0(0,0,s,m_\chi^2,m_t^2,m_\chi^2)
  \ .$$ 
When we come to the contribution from the $SU(2)$ doublet leptoquark $\Omega$,
it is easy to see the simple translation,
$$ \chi^{\fr13} \leftrightarrow  \Omega^{\fr53} \ ,\  
   g^\ell_{L/R} \leftrightarrow  g'^\ell_{L/R} \ ,\ 
   \lambda_\chi \leftrightarrow  \lambda_\Omega \ ,\ 
   m_\chi \leftrightarrow  m_\Omega \ , \  
\hbox{ etc.}    $$
Here $m_\Omega$ is the mass of the $\fr53$ charged leptoquark. More explicitly,
$$ G_\chi=(m_\chi^2+m_t^2) C_0(0,0,s,m_t^2,m_\chi^2,m_t^2)
           +B_0(s,m_t^2,m_t^2)-B_0(0,m_t^2,m_\chi^2)
           + \lambda_\chi v^2 C_0(0,0,s,m_\chi^2,m_t^2,m_\chi^2)
          \ , $$
$$ G_\Omega=(m_\Omega^2+m_t^2) C_0(0,0,s,m_t^2,m_\Omega^2,m_t^2)
           +B_0(s,m_t^2,m_t^2)-B_0(0,m_t^2,m_\Omega^2)
           + \lambda_\Omega v^2 C_0(0,0,s,m_\Omega^2,m_t^2,m_\Omega^2)
           \ . $$
\begin{equation}
\end{equation}
\begin{equation}
 {\cal M}^{\rm ren}(h \to\tau\bar\mu)=-3_c \frac{1}{16\pi^2}\frac{m_t}{v} 
[(G_\chi   g_R^\tau g_L^\mu  + G_\Omega  {g'}_R^\tau {g'}_L^\mu  )L 
+(G_\chi   g_R^\mu  g_L^\tau + G_\Omega  {g'}_R^\mu  {g'}_L^\tau  )R ] 
\ .\end{equation} 
The partial decay width, summing both processes $h\to \tau^\mp\mu^\pm$, is
\begin{equation}
\Gamma(h\to \tau^\mp\mu^\pm)
=
\frac{9}{2048\pi^5} m_h
\left(\frac{m_t}{v}\right)^2
   (|G_\chi   g_R^\tau g_L^\mu  + G_\Omega  {g'}_R^\tau {g'}_L^\mu |^2
   +|G_\chi   g_R^\mu  g_L^\tau + G_\Omega  {g'}_R^\mu {g'}_L^\tau |^2)
\ .\end{equation} 

\section{Physics Possibilities}

\begin{figure}[t!]
\includegraphics[width=3in,angle=-90]{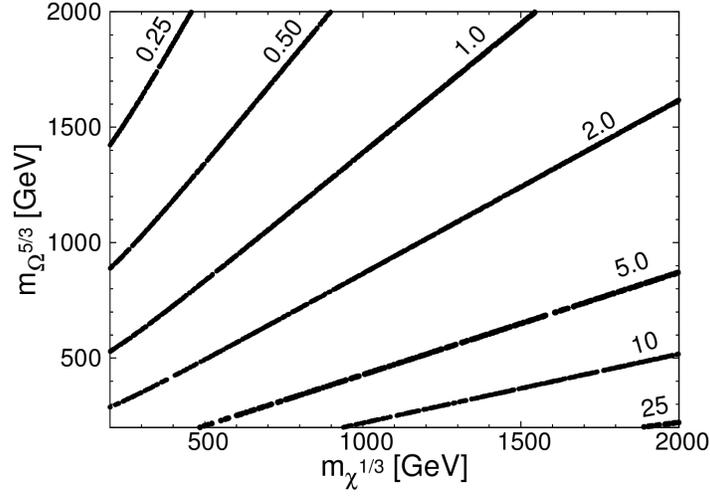}
\caption{\small \label{fig3}
Contour plot of the coupling ratio 
$g_R^\tau  g_L^\mu/({g'}_R^\tau {g'}_L^\mu)$
on the $(m_\chi, m_\Omega)$ plane,
satisfying the tuned cancellation in Eq.~(\ref{can1})
in the amplitude $\tau\to\mu\gamma$. 
}
\end{figure}

To suppress the highly constrained $\tau\to\mu\gamma$, we tune the 
cancellation 
\begin{equation}   
\label{can1}
g_R^\tau  g_L^\mu  x_t H_1(x_t) + {g'}_R^\tau{g'}_L^\mu x'_t H_2(x'_t) 
\approx 0  \ ,  \end{equation}
\begin{equation}
 g_L^\tau  g_R^\mu  x_t H_1(x_t) + {g'}_L^\tau{g'}_R^\mu x'_t H_2(x'_t)
\approx 0   \ .\end{equation}
We choose a simplified scenario that only one chiral mode of the  
muon interactions  is important. Say, $g_L^\mu \gg g_R^\mu$ and 
${g'}_L^\mu \gg {g'}_R^\mu $, then we only finely tune 
the corresponding one constraint, {\it i.e. the first of the two}.
The ratio of the couplings $g_R^\tau  g_L^\mu/({g'}_R^\tau{g'}_L^\mu) $
is given in Fig.~\ref{fig3} for the tuned cancellation in $\tau\to\mu\gamma$.
The contour plot demonstrates that a large parameter space remains available 
for the required fine-tuning.

\begin{figure}[t!]
\includegraphics[width=4in]{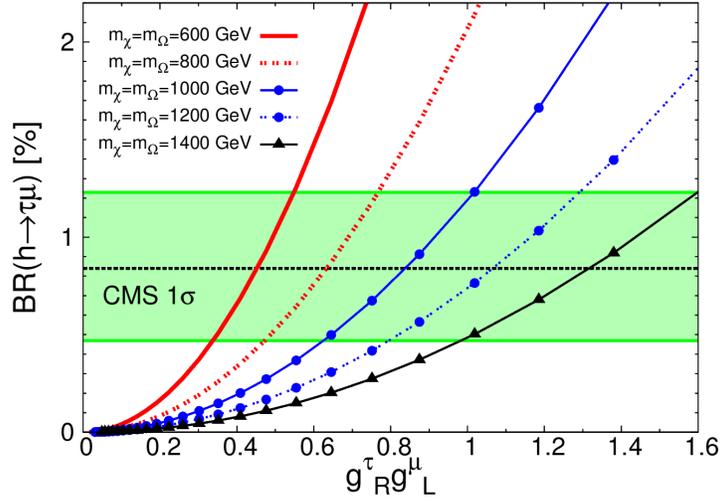}
\caption{\small \label{fig4}
The predicted numerical size of
Br($h\to \tau^\mp\mu^\pm$~both) versus $g_R^\tau  g_L^\mu$ for
various LQ masses when the tuned cancellation is satisfied.
The CMS $1\sigma$ range
of Eq.~(\ref{data}) is also shown.}
\end{figure}

Figure~\ref{fig4} shows the predicted numerical size of 
Br($h\to \tau^\mp\mu^\pm$~both) versus $g_R^\tau  g_L^\mu$ for 
various LQ masses when the tuned cancellation is satisfied.
We have set $\lambda_{\chi,\Omega}=0$ in our numerical study.
A desirable branching fraction at 1\% level occurs for the coupling product
$g_R^\tau  g_L^\mu  \simeq  0.3-1$ for the cases that
$m_\Omega=m_\chi$ from 600 GeV to 1 TeV.

\begin{figure}[t!]
\includegraphics[width=4in]{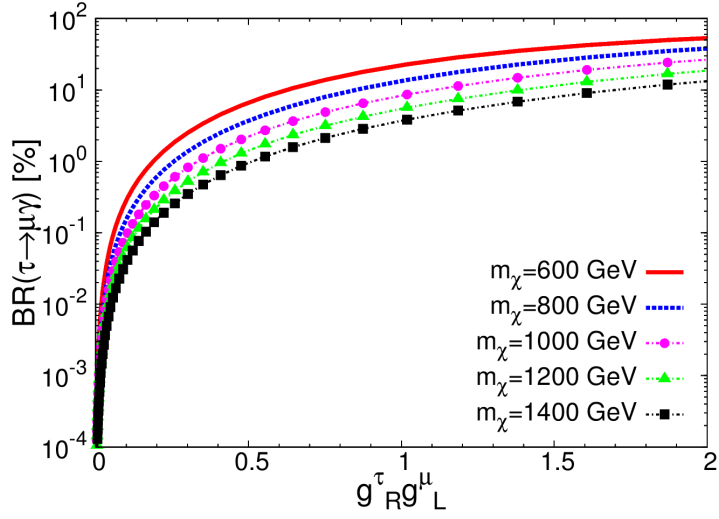}
\caption{\small \label{fig5}
Individual contribution to the Br($\tau\to\mu\gamma)$
if either one of the two canceling amplitudes is switched off.
}
\end{figure}

If we switch off either one of the canceling amplitudes in $\tau\to\mu\gamma$, 
the individual contribution to the Br($\tau\to\mu\gamma)$ is shown in 
Fig.~\ref{fig5}.
This demonstrates how much fine-tuning is required. 
The Br($\tau\to\mu\gamma)$ would be  at about the 1 \% level 
for 500 GeV LQ and $g_R^\tau  g_L^\mu$ about 0.3 to 0.8 if only using 
one of the  two canceling amplitudes.
To go down from $10^{-2}$ to $10^{-8}$ in the branching ratio,
the two amplitudes are required to cancel each other by almost 
one part in $10^3$.

Reference~\cite{Dorsner:2015mja} also proposed a mechanism of 
cancellation in the 
amplitude of $\tau\to \mu\gamma$ with the help of an additional  
vectorial top-like quark. 
However, the detailed gauge quantum numbers of the added structure 
have not been shown to be feasible.

\begin{figure}[t!]
\includegraphics[width=4in]{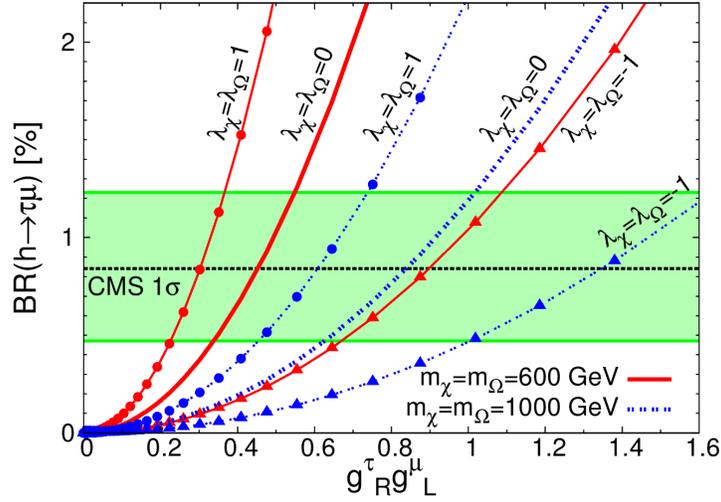}
\caption{\small \label{fig6}
The branching ratio
Br($h\to \tau^\mp\mu^\pm$~both) versus $g_R^\tau  g_L^\mu$ 
for various choices of $\lambda_\chi = \lambda_{\Omega}=-1,0,1$.
The tuned cancellation is satisfied.
}
\end{figure}

So far, we have set $\lambda_{\chi,\Omega}=0$. We show in Fig.~\ref{fig6}
the branching ratio Br$(h \to \tau^\mp \mu^\pm)$ for various 
choices of $\lambda_\chi = \lambda_{\Omega} = -1, 0, 1$.
The tuned cancellation of Eq.~(\ref{can1}) is satisfied.
It gives additional freedom to achieve the desirable branching ratio
for the rare Higgs decay.

\section{Concluding remarks}

The rare decay of $h\to \tau^\mp\mu^\pm$ can be at the current
reachable sensitivity through the LFV LQ interactions, however
fine-tuning is needed to avoid the stringent constraint from the null
observation of $\tau\to\mu\gamma$.
Here we have invoked more than one LQs, which couple to the third
generation quarks, and the second and third generation leptons, in order
to achieve a cancellation in $\tau \to \mu\gamma$ but sizable contributions
to $h \to \tau^\mp \mu^\pm$.

There is another issue related to possible contributions of these LQs to
the muon anomalous magnetic moments (aka $g-2$). It was shown a long time
ago \cite{kingman-lq} and more recently \cite{que-lq}
that if we choose, as we have chosen in the above
analysis, the left-handed coupling to be much larger than the right-handed
coupling for the muon, i.e. $g_L^\mu \gg  g_R^\mu$ and 
${g'}_L^\mu \gg {g'}_R^\mu$,
the LQ contribution to $g-2$ is highly suppressed by $m_\mu/ M_{LQ}$ and
very small for the LQ mass range that we considered in this work.

The required leptoquarks $\chi$ and $\Omega$  
can be strongly produced at the high energy and high luminosity hadron
colliders. They have  dominant decay channels into the top quark and the
charged lepton $\tau$ or $\mu$. That is a very identifiable signature.
Both ATLAS \cite{atlas-lq} and CMS \cite{cms-lq1,cms-lq2} collaborations have 
searched for the third generation leptoquarks via pair production by strong
interaction. 
The CMS have searched for the third generation LQ with electric
charged $-1/3$ (similar to the $\chi^{-1/3}$ of this work) decaying to
a top quark and a tau lepton.  They put a limit of 685 GeV at 95\% CL on 
$m_{\chi}$ \cite{cms-lq1}. On the other hand, both ATLAS \cite{atlas-lq}
and CMS \cite{cms-lq2} searched for the third generation LQ 
with electric charge $-2/3$ decaying into a $\bar b$ antiquark and a tau lepton
(similar to $\Omega^{-2/3}$ in this work), and put a limit of 534 and
740 GeV, respectively.

Therefore, there are still plenty of mass ranges for $\chi^{-1/3}$ beyond
685 GeV and for $\Omega^{-2/3,-5/3}$ beyond 740 GeV that one can directly
search for a top or bottom quark with a tau or muon  at the 
Run 2 of LHC-13.

\section*{\bf Acknowledgments}
We thank N. Kosnik (on behalf of authors of Ref.\cite{Dorsner:2015mja})
for the communication about the  external wave function renomalization in the process
$h\to \tau\mu$.
W.-Y. K. thanks the National Center of Theoretical Sciences and
Academia Sinica, Taiwan, R.O.C. for hospitality.  This research was
supported in parts by the Ministry of Science and Technology (MOST) of
Taiwan under Grant Nos. 102-2112-M-007-015-MY3
and by US DOE under Grant No. DE-FG-02-12ER41811.



\begin{thebibliography}{99}
\bibitem{higgcision}
See for example, 
K.~Cheung, J.~S.~Lee and P.~Y.~Tseng,
  Phys.\ Rev.\ D {\bf 90}, 095009 (2014)
  [arXiv:1407.8236 [hep-ph]].

\bibitem{Khachatryan:2015kon} 
  V.~Khachatryan {\it et al.} [CMS Collaboration],
  arXiv:1502.07400 [hep-ex].
\bibitem{Aubert:2009ag} 
  B.~Aubert {\it et al.} [BaBar Collaboration],
  Phys.\ Rev.\ Lett.\  {\bf 104}, 021802 (2010)
  [arXiv:0908.2381 [hep-ex]].
\bibitem{Dorsner:2015mja} 
  I.~Dorsner, S.~Fajfer, A.~Greljo, J.~F.~Kamenik, N.~Kosnik and 
  I.~Nisandzic,
  JHEP {\bf 1506}, 108 (2015)
  [arXiv:1502.07784 [hep-ph]].
%
\bibitem{Harnik:2012pb} 
  R.~Harnik, J.~Kopp and J.~Zupan,
  JHEP {\bf 1303}, 026 (2013)
  [arXiv:1209.1397 [hep-ph]].
\bibitem{Campos:2014zaa} 
  M.~D.~Campos, A.~E.~C.~Hernández, H.~Päs and E.~Schumacher,
  Phys.\ Rev.\ D {\bf 91}, no. 11, 116011 (2015)
  [arXiv:1408.1652 [hep-ph]].
%
\bibitem{Celis:2014roa} 
  A.~Celis, V.~Cirigliano and E.~Passemar,
  arXiv:1409.4439 [hep-ph].
%
\bibitem{Sierra:2014nqa} 
  D.~Aristizabal Sierra and A.~Vicente,
  Phys.\ Rev.\ D {\bf 90}, no. 11, 115004 (2014)
  [arXiv:1409.7690 [hep-ph]].
\bibitem{Lee:2014rba} 
  C.~J.~Lee and J.~Tandean,
  JHEP {\bf 1504}, 174 (2015)
  [arXiv:1410.6803 [hep-ph]].
\bibitem{Heeck:2014qea} 
  J.~Heeck, M.~Holthausen, W.~Rodejohann and Y.~Shimizu,
  Nucl.\ Phys.\ B {\bf 896}, 281 (2015)
  [arXiv:1412.3671 [hep-ph]].
\bibitem{Crivellin:2015mga} 
  A.~Crivellin, G.~D†¢Ambrosio and J.~Heeck,
  Phys.\ Rev.\ Lett.\  {\bf 114}, 151801 (2015)
  [arXiv:1501.00993 [hep-ph]].
\bibitem{deLima:2015pqa} 
  L.~de Lima, C.~S.~Machado, R.~D.~Matheus and L.~A.~F.~do Prado,
  arXiv:1501.06923 [hep-ph].
\bibitem{11} 
  I.~de Medeiros Varzielas and G.~Hiller,
  JHEP {\bf 1506}, 072 (2015)
  [arXiv:1503.01084 [hep-ph]].
\bibitem{Das:2015zwa} 
  D.~Das and A.~Kundu,
  Phys.\ Rev.\ D {\bf 92}, no. 1, 015009 (2015)
  [arXiv:1504.01125 [hep-ph]].
\bibitem{12}
  A.~Crivellin, G.~D†¢Ambrosio and J.~Heeck,
  Phys.\ Rev.\ D {\bf 91}, no. 7, 075006 (2015)
  [arXiv:1503.03477 [hep-ph]].
\bibitem{13} 
  B.~Bhattacherjee, S.~Chakraborty and S.~Mukherjee,
  arXiv:1505.02688 [hep-ph].
\bibitem{14} 
  Y.~n.~Mao and S.~h.~Zhu,
  arXiv:1505.07668 [hep-ph].
\bibitem{15} 
  B.~Altunkaynak, W.~S.~Hou, C.~Kao, M.~Kohda and B.~McCoy,
  arXiv:1506.00651 [hep-ph].
\bibitem{He:2015rqa} 
  X.~G.~He, J.~Tandean and Y.~J.~Zheng,
  arXiv:1507.02673 [hep-ph].
\bibitem{16} 
  C.~W.~Chiang, H.~Fukuda, M.~Takeuchi and T.~T.~Yanagida,
  arXiv:1507.04354 [hep-ph].
\bibitem{17} 
  W.~Altmannshofer, S.~Gori, A.~L.~Kagan, L.~Silvestrini and J.~Zupan,
  arXiv:1507.07927 [hep-ph].
\bibitem{18}
  A.~Crivellin, J.~Heeck and P.~Stoffer,
  arXiv:1507.07567 [hep-ph].

\bibitem{old}
A.~Pilaftsis,
  Phys.\ Lett.\ B {\bf 285}, 68 (1992);
J.~G.~Korner, A.~Pilaftsis and K.~Schilcher,
  Phys.\ Rev.\ D {\bf 47}, 1080 (1993)
  [hep-ph/9301289].
\bibitem{Lavoura:2003xp} 
  L.~Lavoura,
  Eur.\ Phys.\ J.\ C {\bf 29}, 191 (2003)
  [hep-ph/0302221].
\bibitem{Passarino:1978jh} 
  G.~Passarino and M.~J.~G.~Veltman,
  Nucl.\ Phys.\ B {\bf 160}, 151 (1979).

\bibitem{kingman-lq}
K.~m.~Cheung,
  Phys.\ Rev.\ D {\bf 64}, 033001 (2001)
  [hep-ph/0102238].

\bibitem{que-lq}
F.~S.~Queiroz and W.~Shepherd,
  Phys.\ Rev.\ D {\bf 89}, no. 9, 095024 (2014)
  [arXiv:1403.2309 [hep-ph]].

\bibitem{cms-lq1}
V.~Khachatryan {\it et al.} [CMS Collaboration],
  JHEP {\bf 1507}, 042 (2015)
  [arXiv:1503.09049 [hep-ex]]/

\bibitem{atlas-lq}
G.~Aad {\it et al.} [ATLAS Collaboration],
  JHEP {\bf 1306}, 033 (2013)
  [arXiv:1303.0526 [hep-ex]].

\bibitem{cms-lq2}
V.~Khachatryan {\it et al.} [CMS Collaboration],
  Phys.\ Lett.\ B {\bf 739}, 229 (2014)
  [arXiv:1408.0806 [hep-ex]].

\end{thebibliography}
\end{document}